# Design-Oriented Transient Stability Analysis of Grid-Connected Converters with Power Synchronization Control

Wu, Heng; Wang, Xiongfei

# Design-Oriented Transient Stability Analysis of Grid-Connected Converters with Power Synchronization Control

Heng Wu, *Student Member, IEEE*, and Xiongfei Wang, *Senior Member, IEEE*

*Abstract*—The power synchronization control (PSC) has been increasingly used with voltage-source converters (VSCs) connected to the weak ac grid. This paper presents an in-depth analysis on the transient stability of the PSC-VSC by means of the phase portrait. It is revealed that the PSC-VSC will maintain synchronization with the grid as long as there are equilibrium points after the transient disturbance. In contrast, during grid faults without any equilibrium points, the critical clearing angle (CCA) for the PSC-VSC is identified, which is found equal to the power angle at the unstable equilibrium point of the post-fault operation. This fixed CCA facilitates the design of power system protection. Moreover, it is also found that the PSC-VSC can still re-synchronize with the grid after around one cycle of oscillation, even if the fault-clearing angle is beyond the CCA. This feature reduces the risk of system collapse caused by the delayed fault clearance. These findings are corroborated by simulations and experimental tests.

## I. INTRODUCTION

THE increasing use of power electronic converters poses new challenges to the stability and control of modern electrical energy systems. Although the vector current control is still the dominant control scheme for voltage source converters (VSCs) in different applications [1]-[2], it has some limitations when the VSC is connected to a very weak grid. More specifically, it is difficult to guarantee the stability of the VSC with the vector current control when the short circuit ratio (SCR) of the ac grid is approaching 1.3, due to the negative impact of the phase locked loop (PLL) [3]-[4].

To tackle this stability challenge, the concept of power synchronization control (PSC) is developed [5]. This control strategy enables the VSC to operate with ultra-weak ac grid, where the SCR is nearly equal to 1 [6]. The PSC method has been wildly adopted in different applications, such as the high-voltage direct-current (HVDC) transmission systems [7]-[9], photovoltaic inverters [10], and grid-side converters of wind turbines [11].

In recent years, the small-signal stability of the PSC-VSC has been extensively studied by means of the Jacobian transfer matrix [12] or impedance models [13]-[14]. It is revealed that the PSC exhibits more robust small-signal dynamics than that of the vector current control for VSCs operating in weak grid conditions [15]. In contrast to the small-signal stability study, less research works are found on the transient stability study of the PSC-VSC, i.e., the ability of VSCs to maintain synchronization with the power system when subjected to transient disturbances, e.g., the faults on power transmission lines, the loss of generators, or the large load swing [16]. Although abundant research results are available for the transient stability study of synchronous generators (SGs) [16], they cannot be directly applied to the PSC-VSC due to their different dynamics. The PSC is actually the first-order control system [5], which is different from the second-order swing equation of SGs.

The major difficulty for the transient stability analysis of the PSC-VSC lies in extracting its large-signal nonlinear dynamic responses, where the small-signal modeling method and the linear control theory do not apply, and it is usually a challenging task to perform the theoretical analysis based on nonlinear differential equations [17]. Hence, the time-domain simulation is still the dominant tool for the transient stability analysis [16]. Although the time-domain simulation can yield accurate results, it is often time consuming and provides little insight into the underlying transient stability mechanism of the PSC-VSC. Hence, a design-oriented transient stability analysis that can characterize the dynamic impact of the PSC is urgently demanded.

To the best knowledge of authors, the transient stability of the PSC-VSC is hitherto unaddressed in the literature. This paper attempts to fill this void. First, a conceptual review of the transient stability problem in traditional SG-based power systems is presented, which intends to summarize the basics and critical parameters of the transient stability study. Then, the dynamic model of the PSC-VSC for the transient stability analysis is developed, and the phase portrait is employed to characterize the transient behavior of the PSC-VSC based on a single-converter infinite-bus system. It is revealed that the PSC-VSC exhibits superior transient stability performance than that of SGs. Moreover, during grid faults where the system has no equilibrium points, the critical clearing angle (CCA) and the critical clearing time (CCT), which are the maximum fault clearing angle and time for guaranteeing the system stability, are analytically derived, and are found as fixed values corresponding to the unstable equilibrium point (UEP) during the post-fault operation. Such a fixed CCA/CCT facilitates the design of power system protection. Further, it is also found that the PSC-VSC is able to re-synchronize with the grid after around one cycle of oscillation, even if the fault clearing angle/time is beyond the CCA/CCT. This self-restoration property for the PSC-VSC reduces the risk of system collapse caused by the delayed fault clearance. Finally, these theoretical findings are verified by time-domain simulations and experimental tests. It is worth noting that the findings in this paper are not only limited to the PSC-VSC, but equally apply to the droop-control-based VSC, due to the equivalence between the PSC and the droop control [18]. Moreover, the results can also be extended to any other control schemes with the first-order synchronization loop.

## II. TRANSIENT STABILITY BASICS OF SG-BASED POWER SYSTEMS

Fig. 1 illustrates a general per-phase diagram of a single-machine infinite-bus system, where the SG feeds power to the grid through two paralleled transmission lines. S1~S4 are circuit breakers, and $X_{gnd}$ represents the grounding impedance when the symmetrical three-phase to ground fault occurs. The swing equation of the SG can be expressed as [16]

$$P_m - P_e - D\dot{\delta} = J\omega_n \ddot{\delta} \qquad (1)$$

where $P_m$ and $P_e$ are the input mechanical power and the output electrical power of the SG, respectively. $J$ denotes the inertia. $D$ is the damping coefficient, and its unit is W·s/rad. $\omega_n$ is the rated angular speed of the rotor.

The output power of the SG is given by [16]

$$P_e = \frac{3V_{PCC}V_g}{2X_g}\sin\delta \qquad (2)$$

where $V_{PCC}$ and $V_g$ are magnitudes of the SG terminal voltage and the infinite bus voltage, respectively. $X_g$ is the equivalent impedance between the SG and the infinite bus, and $\delta$ is the power angle, i.e., the phase difference between the voltage at the point of common coupling (PCC) and the infinite bus voltage. Based on (2), $P_e$-$\delta$ curves with both two transmission lines in service can be plotted, which are shown as dashed lines in Fig. 2(a) and (b). Obviously, the SG initially operates at the equilibrium point $a$, where $P_m = P_e$ at the steady state.

There are, in general, two types of transient stability problems, depending on whether or not the system has equilibrium points during a disturbance, which are:

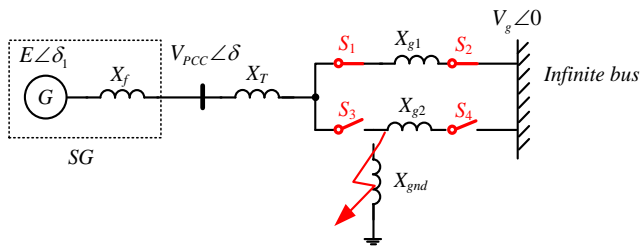

Fig. 1. The SG connecting to the infinite bus.

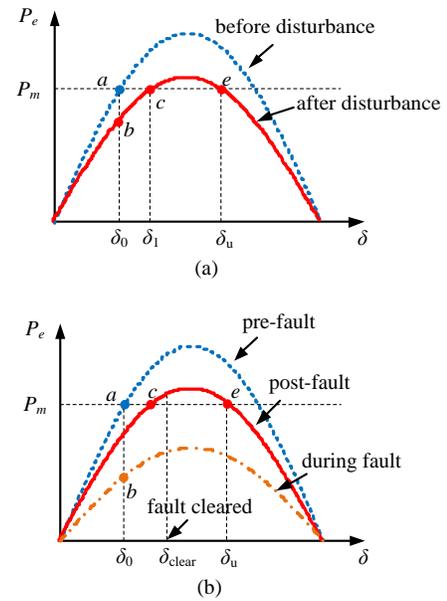

Fig. 2. $P_e$-$\delta$ curves. (a) With equilibrium points after the transient disturbance, dashed line: before disturbance, solid line: after disturbance. (b) Without equilibrium points during the transient disturbance, dashed line: pre-fault, dashed dotted line: during fault, solid line: post-fault.

### A. Type-I: Presence of Equilibrium Points during a Disturbance

First, for the type-I transient stability problem, there are equilibrium points after the disturbance. As shown in Fig. 1, the disturbance is considered as a sudden disconnection of one of the transmission lines (e.g., $X_{g2}$), which leads to a higher $X_g$, and the resulting $P_e$-$\delta$ curve is shown as the solid line in Fig. 2(a). The point $c$ and $e$ represent the stable equilibrium point (SEP) and the UEP after the disturbance, respectively [16]. According to the swing equation given by (1), the rotor of the SG accelerates from the point $b$ till the point $c$, due to $P_m > P_e$, and decelerates thereafter. Yet, the power angle $\delta$ keeps increasing whenever the rotor speed of the SG is higher than the synchronous speed. Hence, the system will be kept stable if and only if the rotor speed is recovered to the synchronous speed before the UEP, i.e., the point $e$ [16]. Otherwise, the rotor accelerates again after the point $e$, due to $P_m > P_e$, and the SG finally loses the synchronization with the power grid.

### B. Type-II: No Equilibrium Points during a Disturbance

Second, for the type-II transient stability problem, there are no equilibrium points during the disturbance. A symmetrical three-phase to ground fault on $X_{g2}$ is considered as the transient disturbance in this case, and the resulting $P_e$-$\delta$ curve during the fault is shown as the dashed-dotted line in Fig. 2(b). Due to the absence of equilibrium points, $P_m > P_e$ always holds. It is known from (1) that the rotor of the SG keeps accelerating. To prevent the system from collapsing, the protective relay has to be activated to clear the fault by opening the circuit breakers S3 and S4. Consequently, the post-fault $P_e$-$\delta$ curve is shown as the solid line in Fig. 2(b). The rotor of the SG decelerates after the fault is cleared and the system will be stable if its rotor speed can be recovered to the synchronous speed before the UEP (the point $e$). Hence, the fault clearing becomes crucial to restore the system with equilibrium points, and the CCA and the CCT are critical parameters for guaranteeing the transient stability of the power system.

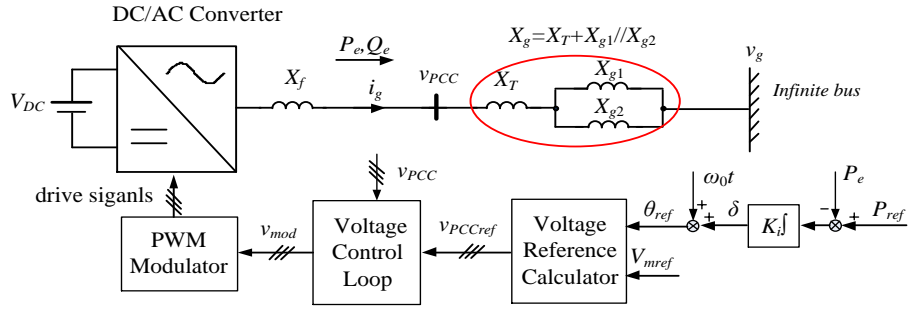

Fig. 3. Control diagram of the PSC-VSC under the normal operation, where its output current is within the current limit.

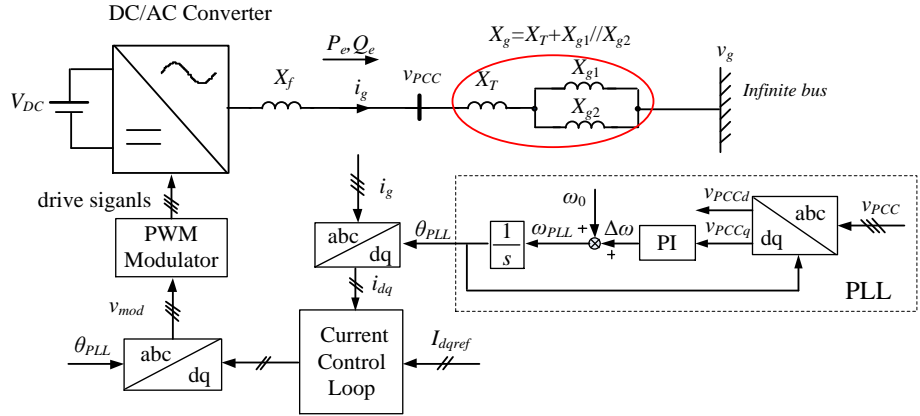

Fig. 4. The PSC-VSC is switched to the vector current control when its output current reaches the current limit.

## III. TRANSIENT STABILITY ANALYSIS OF THE PSC-VSC

### A. Dynamic Representation of the PSC-VSC

Fig. 3 shows the one-line diagram of a three-phase PSC-VSC with the alternating-voltage control (AVC) [5]. A constant dc-link voltage ($V_{dc}$) of the VSC is assumed. The PSC-VSC is connected to the infinite bus through two paralleled transmission lines $X_{g1}$ and $X_{g2}$. In this work, the transmission line is represented by its inductance and its parasitic shunt capacitance is not considered, due to the fact that the shunt capacitance has little impact on the transient stability of the system [16]. $v_g$ denotes the voltage of the infinite bus while $i_g$ denotes the current of the PSC-VSC injected into the infinite bus. $X_g$ and $X_f$ represent the equivalent grid impedance and the impedance of the converter output filter, respectively. An inner voltage loop is used to regulate the voltage at the point of common coupling ($v_{PCC}$) to track its reference $v_{PCCref}$, and the overcurrent limitation is embedded in the voltage loop [5]. $V_{mref}$ and $\theta_{ref}$ denote the magnitude and the phase angle of $v_{PCCref}$, respectively. In the AVC mode, the control system of the VSC regulates its active power injected to the ac grid while keeping the PCC voltage constant, and the injected reactive power is not directly controlled [5]. Therefore, $V_{mref}$ is usually set as a constant value, e.g., $V_{mref}$ =1.0 p.u., while $\theta_{ref}$ is generated by the PSC, which is given by:

$$\theta_{ref} = K_i \int \left( P_{ref} - P_e \right) + \omega_0 t \quad (3)$$

where $P_{ref}$ is the active power reference for the PSC-VSC, which is decided by the power system operator [5]. $P_e$ is the output active power of the PSC-VSC, and its expression is given by (2). $\omega_0$ denotes the grid frequency and $K_i$ is the integral gain of the PSC.

It is worth noting that the timescale of the inner voltage loop is usually around 1 ms - 10 ms [19], whereas the timescale of the outer power loop (PSC loop) is usually 10 times higher, i.e., around 100ms - 1s [19]. Hence, the dynamics of these two loops can be analyzed individually, due to their decoupled timescales [20]. Therefore, when analyzing the transient stability impact of the PSC loop, $v_{PCC}$ can be considered to track its reference value instantaneously, i.e., $v_{PCC}=v_{PCCref}$ [20].

On the other hand, for large grid disturbances that trigger the overcurrent limit of the PSC-VSC, e.g., the low-impedance fault or the short circuit fault [21], the PSC-VSC is switched to the vector current control [5], as shown in Fig. 4. $I_{dqref}$ and $i_{dq}$ represent the current references and grid currents in the $dq$ frame, respectively. The current loop is used to regulate the output current of the VSC to track its reference in order to avoid the overcurrent. The synchronization between the VSC and the power grid is realized by the PLL, and the grid synchronization mechanism of the PLL can be found in [22]. In this case, transient dynamics of the VSC is determined by the PLL [23], which are different from the PSC. The transient stability of the VSC with the vector current control and the influence of the PLL has already been discussed in [23]-[26], and is out of scope of this work. Hence, only the transient disturbances that do not trigger the overcurrent limit of the PSC-VSC, e.g., the sudden disconnection of one of transmission lines and the high impedance fault [27], are considered. In time-domain simulations and experiments of this work, the overcurrent limit is set as 1.8 p.u.

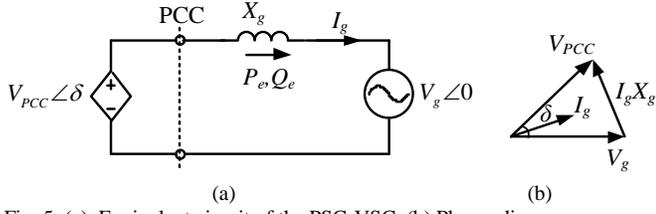

Fig. 5. (a) Equivalent circuit of the PSC-VSC. (b) Phasor diagram.

Fig. 5 (a) shows the equivalent circuit of the PSC-VSC, while its phasor diagram is given in Fig. 5 (b). $\delta$ denotes the power angle, which represents the phase difference between the PCC voltage and the infinite bus voltage. The initial phase angle of the infinite bus voltage is defined as zero, i.e., $\theta_g = \omega_0 t$. Therefore, $\delta$ can be expressed as follow based on (3)

$$\delta = \theta_{ref} - \theta_g = K_i \int (P_{ref} - P_e) + \omega_0 t - \omega_0 t \\ = K_i \int (P_{ref} - P_e) \quad (4)$$

Substituting (2) into (4) and considering the assumption of $V_{PCC} = V_{mref}$, yielding

$$\dot{\delta} = K_i \left( P_{ref} - \frac{3V_{PCC}V_g}{2X_g} \sin \delta \right) \\ = K_i \left( P_{ref} - \frac{3V_{mref}V_g}{2X_g} \sin \delta \right) \quad (5)$$

From (5), it is obvious that the PSC loop is represented by the first-order nonlinear dynamic equation, which is different from the second-order swing equation of the SG given in (1). Hence, the methods and findings of the transient stability assessment for the SG-based power system cannot be extended to the PSC-VSC.

### B. Phase Portrait of the PSC-VSC

Basically, the transient stability of the PSC-VSC is related to the dynamic response of the power angle $\delta$ after the large disturbance. The system will be stable if $\delta$ is converged to the new equilibrium value as $t \to \infty$, and will be unstable if it is diverged. The analytical expression of $\delta$ can be obtained by solving (5).

Eq. (5) can be rewritten as

$$dt = \frac{d\delta}{a - b \sin \delta} \quad (6)$$

where

$$a = K_i P_{ref} \qquad b = K_i \frac{3V_{mref}V_g}{2X_g} \quad (7)$$

Integrating both sides of (6), yielding

$$t = \int dt = \int \frac{d\delta}{a - b \sin \delta} \quad (8)$$

The complete mathematical deduction procedure for solving (8) is provided in the Appendix, and only the result is given here

$$t = \frac{2}{\sqrt{a^2 - b^2}} \arctan \left[ \frac{\tan(\delta/2) - b/a}{\sqrt{1 - (b/a)^2}} \right] + C \quad (9)$$

where $C$ is the constant determined by the initial condition. Assuming $\delta = \delta_0$ at $t = 0$, (9) can be rewritten as

$$t = \frac{2}{\sqrt{a^2 - b^2}} \arctan \left[ \frac{\tan(\delta/2) - b/a}{\sqrt{1 - (b/a)^2}} \right] \\ - \frac{2}{\sqrt{a^2 - b^2}} \arctan \left[ \frac{\tan(\delta_0/2) - b/a}{\sqrt{1 - (b/a)^2}} \right] \quad (10)$$

While the result of (10) is accurate, it provides little physical insight on the dynamic response of $\delta$ under the given initial condition. Therefore, it is not intuitive to analyze the transient behavior of the PSC-VSC directly based on (10).

In contrast, the phase portrait, which is a graphical solution of (5), is capable of providing a simpler and more intuitive result. Instead of solving $\delta(t)$ analytically, the $\dot{\delta} - \delta$ curve can be readily plotted based on the function defined by (5), and there are three different scenarios based on whether the system has equilibrium points, as shown in Fig. 6. These curves are the so-called phase portraits [17]. At each point, the changing trend of $\delta$ is determined by its derivative $\dot{\delta}$, e.g., $\delta$ will increase if $\dot{\delta} > 0$ and decrease if $\dot{\delta} < 0$, which are indicated by arrows in

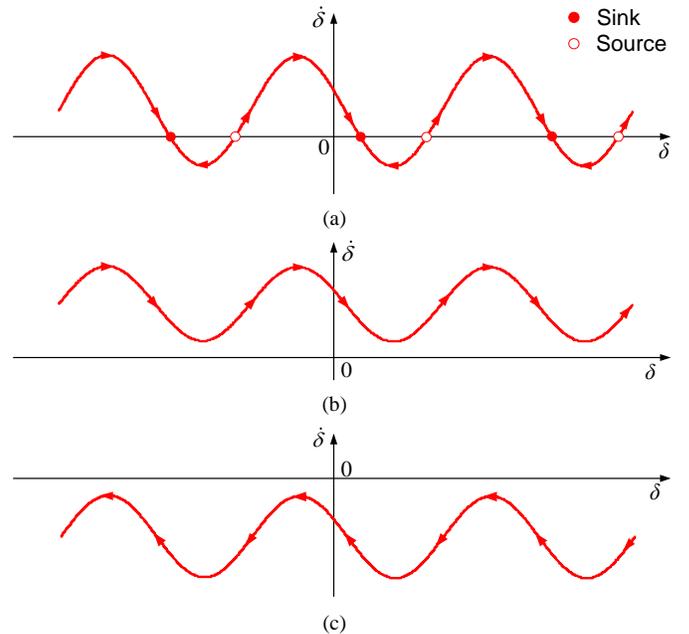

Fig. 6. Phase portraits of the PSC-VSC (a) With equilibrium points. (b) Without equilibrium points, $\dot{\delta}$ is always larger than zero. (c) Without equilibrium points, $\dot{\delta}$ is always smaller than zero.

Fig. 6. The points at which $\dot{\delta}=0$ are the equilibrium points. As shown in Fig. 6 (a), the solid dots represent the SEPs (also called sinks), while the open circles represent the UEPs (also called sources) [17]. Obviously, for any given initial conditions, the system will either monotonically approach to its nearest SEP if it exists (as shown in Fig. 6(a)), or monotonically move toward $\pm\infty$, if there are no equilibrium points, as shown in Fig. 6(b) and (c). It is important to note that such dynamic behavior is not limited to the PSC-VSC, but is equally applicable to other first-order nonlinear systems [17].

### C. Type-I Transient Stability of the PSC-VSC

In this case, the transient disturbance is considered as the sudden disconnection of the transmission line 2 ($X_{g2}$ in Fig. 3) due to the symmetrical three-phase open-circuit fault, which is same as the case shown in Fig. 2(a), and the system has equilibrium points after the disturbance. The main system parameters used in the analysis are summarized as the test case I in Table I.

Fig. 7 plots the phase portraits based on (5), where the dashed line represents the phase portrait before the disturbance and the solid line is the phase portrait after the disturbance. It is clear that the system trajectory can be readily obtained from the phase portrait, which is highlighted by the solid line with arrows in Fig. 7. Before the disturbance, the system operates at the point $a$ ($\dot{\delta}=0$), which suddenly moves to the point $b$ during the disturbance. Then, the power angle $\delta$ starts to increase, due to $\dot{\delta}>0$, and stops at the SEP $c$ ($\dot{\delta}=0$), where the system comes into a new steady state.

In contrast to Fig. 2(a), the transient response of the PSC-VSC is significantly different from that of the SG. Since $\dot{\delta}$ is equal to zero at the point $c$ for the PSC-VSC, the system will stay at the SEP $c$ once it reaches there, and will not crossover. This fact implies a transient response with no overshoots and oscillations, and thus the PSC-VSC has no transient stability problems, provided that the system has equilibrium points after the disturbance.

### D. Type-II Transient Stability of the PSC-VSC

In the Type-II transient stability study, the considered transient disturbance for the PSC-VSC is same as the case given in Fig. 2(b), i.e., a symmetrical three-phase to ground fault on the transmission line 2 with the high fault impedance, and the VSC has no equilibrium points during the fault. The fault is then cleared by opening the circuit breaker in a certain period of time. The main parameters used in the analysis are summarized as the test case II in Table I.

Fig. 8 plots the phase portraits based on (5), where the dashed line, the dashed-dotted line, and the solid line represent the phase portrait pre-, during and post-fault, respectively. Two cases with different fault clearing angles are shown, and the solid lines with arrows represent the system trajectories in both cases. From Fig. 8(a), it is clear that as long as the fault clearing angle is smaller than $\delta_u$, which is the power angle corresponding to the UEP (point $e$), the post-fault operating point of the system can always move back to the SEP $c$, and the system is kept stable. Hence, $\delta_u$ is identified as the CCA for the PSC-VSC, which can be calculated as:

TABLE I
PARAMETERS FOR TRANSIENT DISTURBANCE TEST (SIMULATION)

| Parameters | | | |
|---|---|---|---|
| $P_{ref}$ | 1000 MW (1 p.u.) | $L_f$ | 0.035 H (0.075 p.u.) |
| $V_g$ | 220 kV/50 Hz (1 p.u.) | $K_i$ | $9.3 \cdot 10^{-9}$ (0.01 p.u.) |
| Test case I | | Test case II | |
| $L_T$ | 0.01 H (0.02 p.u.) | $L_T$ | 0.37 H (0.8 p.u.) |
| $L_{g1}$ | 0.39 H (0.85 p.u.) | $L_{g1}$ | 0.07 H (0.15 p.u.) |
| $L_{g2}$ | 0.39 H (0.85 p.u.) | $L_{g2}$ | 0.37 H (0.8 p.u.) |
| | | $L_{gnd}$ | 0.23 H (0.5 p.u.) |

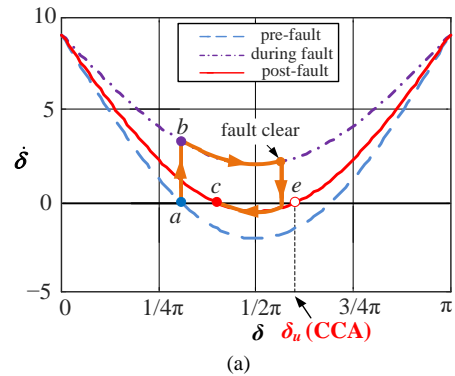

(a)

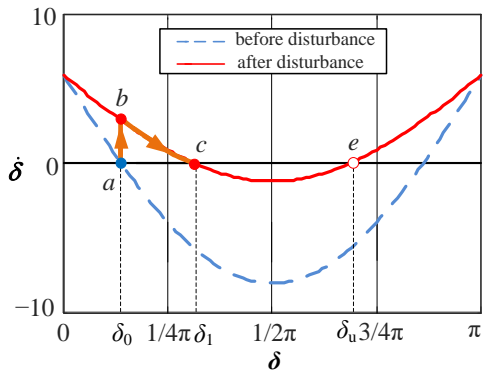

Fig. 7. Phase portraits of the PSC-VSC before (dashed line) and after (solid line) the transient disturbance for the Type-I transient stability.

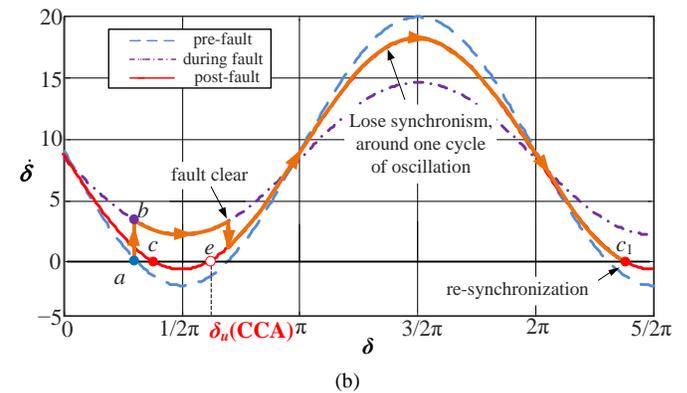

(b)

Fig. 8. Phase portraits of the PSC-VSC before (dashed line), during (dashed dotted line) and post (solid line) the fault. (a) Fault clearing angle lower than $\delta_u$. (b) Fault clearing angle higher than $\delta_u$.

$$\text{CCA} = \delta_u = \pi - \arcsin\frac{2P_{ref}X_g}{3V_{mref}V_g} \quad (11)$$

Obviously, $\delta_u$ is solely determined by the post-fault system configuration and is not affected by different fault conditions. This constant CCA simplifies the design of the protective relay for the PSC-VSC-based power systems.

Moreover, differing from the SG, which may lose synchronization with the power grid if the fault clearing angle is higher than the CCA [16], the PSC-VSC has an attractive feature of self-restoration, i.e., re-synchronizing with the grid after around one cycle of oscillation. The underlying reason for this behavior has already been given in part B of this section, i.e., as long as the system is restored with equilibrium points by clearing the fault (no matter the fault clearing angle is within or beyond the CCA), it will finally stay at the nearest SEP. Fig. 8(b) illustrates the system operating trajectory when the fault is cleared beyond the CCA, where the system finally stays at the new SEP $c_1$ after around one cycle of oscillation. This superior feature prevents the system collapse due to the delayed fault clearance.

The CCT is the corresponding time interval during which the power angle of the PSC-VSC moves from $\delta_0$ to the CCA [16]. By substituting $\delta$=CCA into (10), the CCT can be expressed as

$$CCT = \frac{2}{\sqrt{a^2-b^2}}\arctan\left[\frac{\tan(CCA/2)-b/a}{\sqrt{1-(b/a)^2}}\right]$$
$$\quad - \frac{2}{\sqrt{a^2-b^2}}\arctan\left[\frac{\tan(\delta_0/2)-b/a}{\sqrt{1-(b/a)^2}}\right] \quad (12)$$

Table II summarizes the transient stability of the PSC-VSC, it is clear that its transient behavior depends on the operating scenarios during the fault (with or without equilibrium points, whether or not the current limitation is triggered), rather than the specific type of fault. Therefore, the conclusion in this paper is equally applicable to other kinds of fault, e.g., generator outage, step of load, etc.

Moreover, it is worth mentioning that the transient dynamic behavior of the PSC-VSC is determined by its inherent first-order nonlinear dynamic equation [17]. Hence, the findings and the used analytical approach can be extended to other control schemes with the first-order nonlinear dynamic responses. Yet, the same conclusion does not apply to the control system with a second-order nonlinear dynamic responses, e.g., the virtual synchronous generator control, which mimics the swing equation of synchronous generators [16], or the vector current control using a second-order PLL for synchronizing with the power grid [22]. It has recently been found that using the synchronous reference frame PLL (SRF-PLL) makes the vector current control exhibit a second-order nonlinear dynamic response, which is similar to the swing equation of the SG [23]-[26]. The different system order of the PSC-VSC and the vector current control results in the different transient stability performance. The systematic analysis on the transient stability of the vector current control with the SRF-PLL can be found in [26].

## IV. SIMULATION AND EXPERIMENTAL RESULTS

Time-domain simulations and experimental tests are performed to verify the effectiveness of the transient stability analysis based on the phase portrait. The same transient disturbances considered in Section II and III are used. It is worth noting that the inner voltage loop and the overcurrent limitation are both implemented in simulations and experiments. As will be shown in the following, the simulations and experimental results match closely with the theoretical analysis, which justify the assumptions made in deriving the dynamic representation of the PSC-VSC, i.e., the nonlinear differential equation given by (5).

### A. Simulation Results
*1) Type I transient stability*

Fig. 9 shows the simulation results of a sudden disconnection of the transmission line 2, where parameters of the test case I listed in Table I are used. The system has the equilibrium points after the disturbance. The corresponding power angles of SEPs before and after the transient disturbance are 26.4° and 60.5°, respectively. It can be observed that the VSC smoothly reaches to the new SEP after the transient disturbance without any overshoots or oscillations, and the typical overdamped response of the power angle can be clearly identified. Therefore, the transient instability does not occur for the PSC-VSC, as long as the system has equilibrium points after the disturbance. The simulation results agree well with the phase portrait analysis presented in Fig. 7. It is also noted that the grid current $i_g$ settles with a little bit higher value after the disturbance, it is because of the increased output reactive power of the PSC-VSC after the disturbance, which is resulted from the increased power angle [16].

TABLE II
TRANSIENT STABILITY OF THE PSC-VSC

| Operating scenarios during the transient disturbance | | Transient stability of the PSC-VSC |
|---|---|---|
| With equilibrium points | | No transient stability problem. |
| No equilibrium points | Do not trigger the current limitation. | • CCA and CCT are fixed, and can be calculated based on (11) and (12)<br>• Able to re-synchronize with the power grid even if the fault is cleared beyond the CCT |
| | Trigger the current limitation. | • Switch to the vector current control [5], and the transient stability is mainly determined by the PLL [23]-[26]. |

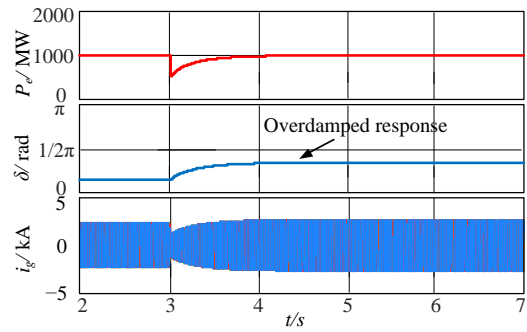

Fig. 9. Simulated dynamic responses of the PSC-VSC after the transient disturbance, where parameters of the test case I listed in Table I are used.

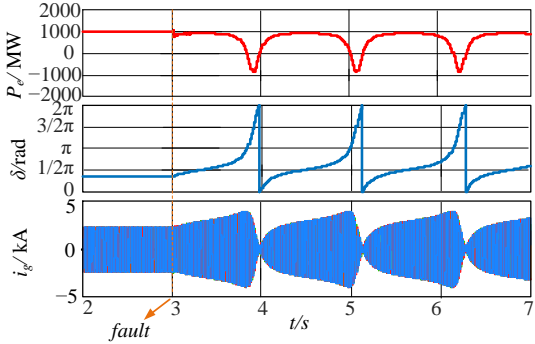

(a)

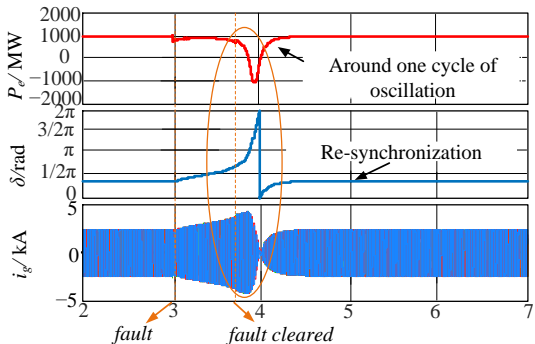

(b)

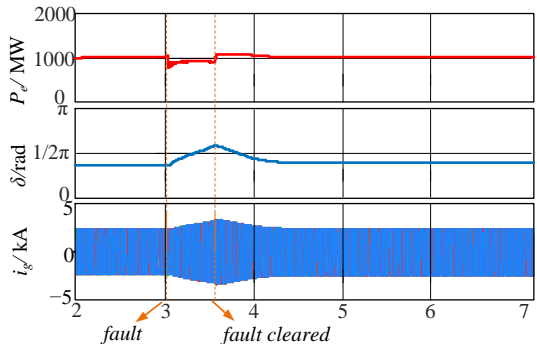

(c)

Fig. 10. Simulated dynamic responses of the PSC-VSC after the transient disturbance, where parameters of the test case II listed in Table I are used. (a) Fault is not cleared. (b) Fault is cleared with the fault clearing time 0.7s > CCT. (c) Fault is cleared with the fault clearing time 0.5s < CCT.

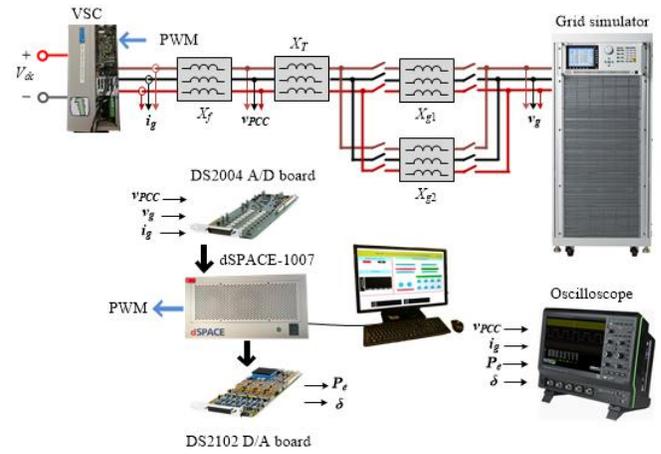

Fig. 11. Configuration of the experimental setup.

### 2) Type II transient stability

Fig. 10 shows the simulation results of the symmetrical three-phase to ground high impedance fault on the transmission line 2, where parameters of the test case II given in Table I are used. Based on (11) and (12), it can be calculated that the CCA and the CCT for the PSC-VSC are 108º, 0.58s, respectively. Three different cases are studied in the simulation, i.e., 1) the fault is not cleared, 2) the fault is cleared with the fault clearing time beyond the CCT, and 3) the fault is cleared with the fault clearing time smaller than the CCT. Since the VSC has no equilibrium points during the fault, it will become unstable if the fault is not cleared, as shown in Fig. 10 (a). The oscillation in the output power, power angle and grid currents can be observed, indicating that the VSC loses synchronization with the power grid. In the case that the fault is cleared but the fault clearing time is beyond the CCT, the VSC automatically re-synchronizes with the power grid after around one cycle of oscillation, as illustrated in Fig. 10 (b), where the fault clearing time is set as 0.7s in the simulation. In contrast, the system can be kept stable if the fault is cleared and the fault clearing time is smaller than the CCT, as illustrated in Fig. 10 (c), where the fault clearing time is set as 0.5s. These simulation results corroborate the phase portrait analysis shown in Fig. 8.

### B. Experimental Results

To further verify the simulation results, the experiments are carried out with a three-phase grid-connected converter with downscaled voltage and power ratings, the per unit values of parameters used in the experiment is same as that used in the simulation, as shown in Table III. Same transient disturbances described in the simulations are considered in the experiments. The experimental setup is shown in Fig. 11, where the three phase inductors are used to represent the line impedance. The control algorithm is implemented in the DS1007 dSPACE system, where the DS2004 high-speed A/D board is used for the voltage and current measurements. The phase angle of the PCC voltage and the grid voltage is detected by the PLL. The power angle is calculated based on $\delta = \theta_{ref} - \theta_g$, and is outputted through the DS2102 high-speed D/A board. The output voltage of the D/A board ranges from 0V to 10V, which corresponds to the power angle ranges from 0 to $2\pi$. A constant dc voltage supply is used at the dc-side, and a 45 kVA Chroma 61850 grid simulator is used to generate the grid voltage.

Fig. 12 shows the experimental results of a sudden disconnection of the transmission line 2, and the system has equilibrium points after the disturbance, the tested overdamped response of the PSC-VSC can be observed. It is obvious that no transient stability problem occurs in this case, and it further confirms the simulation result depicted in Fig. 9.

Fig. 13 shows the experimental results of the symmetrical three-phase to ground high impedance fault on the transmission line 2, and the system does not have equilibrium points during the fault. Based on parameters of the test case II in Table III, the CCA and the CCT for the PSC-VSC are calculated as 108°, 0.58s, respectively. Three different cases are studied in the experiment, i.e., fault is not cleared, fault is cleared with the fault clearing time beyond the CCT, fault is cleared with the fault clearing time smaller than the CCT. Fig. 13(a) shows the measured responses when the fault is not cleared, where the VSC loses synchronism with the power grid. Fig. 13(b) shows the measured responses when the fault is cleared with the fault clearing time 0.7s, which is beyond the CCT. It can be seen that the VSC re-synchronizes with the power grid after around one cycle of oscillation. In contrast, the system is kept stable when the fault is cleared with the fault clearing time smaller than the CCT, as shown in Fig. 13(c), where the fault clearing time is set as 0.5s in the experiment. Obviously, all the experimental results are in accordance with the theoretical analysis and simulation results.

TABLE III
PARAMETERS FOR TRANSIENT DISTURBANCE TEST (EXPERIMENTS)

| Parameters | | | |
|---|---|---|---|
| $P_{ref}$ | 1 kW (1 p.u.) | $L_f$ | 1.5 mH (0.075 p.u.) |
| $V_g$ | 50 V/50 Hz (1 p.u.) | $K_i$ | $9.3 \cdot 10^{-3}$ (0.01 p.u.) |
| Test case I | | Test case II | |
| $L_T$ | 0.5 mH (0.02 p.u.) | $L_T$ | 19 mH (0.8 p.u.) |
| $L_{g1}$ | 21 mH (0.85 p.u.) | $L_{g1}$ | 3 mH (0.15 p.u.) |
| $L_{g2}$ | 21 mH (0.85 p.u.) | $L_{g2}$ | 19 mH (0.8 p.u.) |
| | | $L_{gnd}$ | 12 mH (0.5 p.u.) |

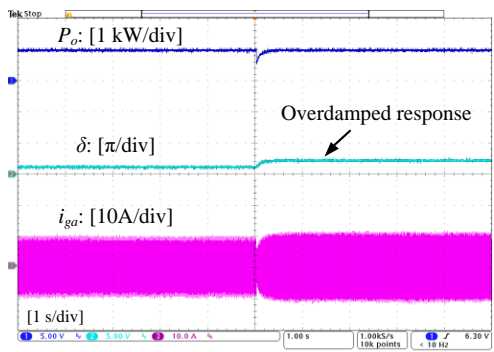

Fig.12. Experimental results of the dynamic response of the PSC-VSC after the transient disturbance, where parameters of the test case I listed in Table III are used.

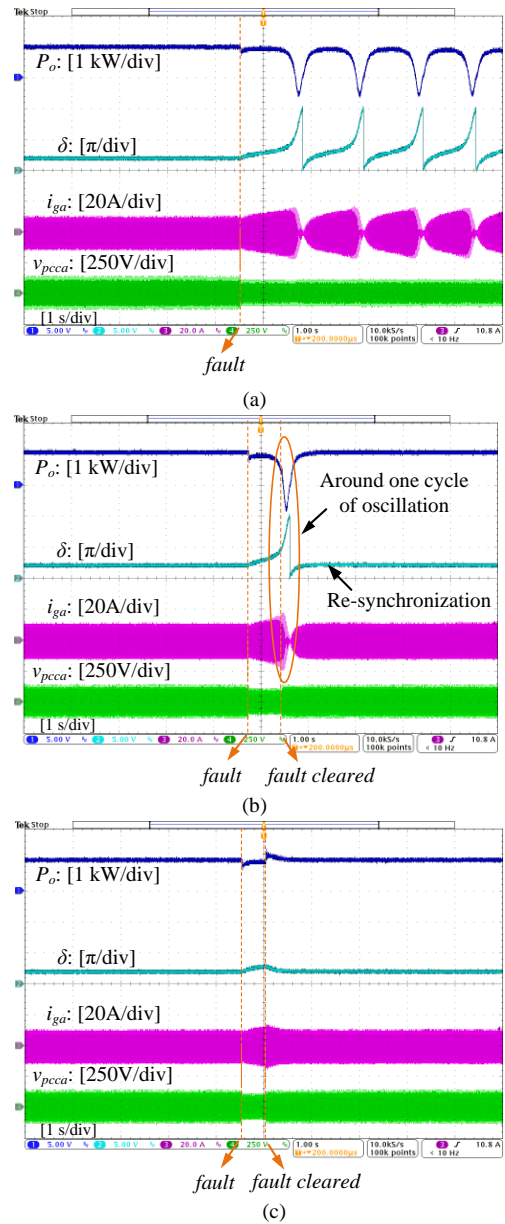

Fig. 13. Experimental results of dynamic responses of the PSC-VSC after the transient disturbance, where parameters of the test case II listed in Table III are used. (a) Fault is not cleared. (b) Fault is cleared with fault clearing time 0.7s > CCT. (c) Fault is cleared with fault clearing time 0.5s < CCT.

## V. CONCLUSION

In this paper, the transient stability of the PSC-VSC has been discussed in the single-converter infinite-bus scenario. The transient responses of the PSC-VSC have been elaborated by means of the phase portrait. The first-order power synchronization loop brings significantly different transient dynamics of the PSC-VSC, compared to the traditional SG-based power systems. The major findings of the paper are summarized as follows:

1) In the case that the system has equilibrium points during the disturbance, the PSC-VSC has no transient stability problem, due to its overdamped dynamic response.

2) For the system that has no equilibrium points during the disturbance, the CCA of the PSC-VSC is identified as the power

angle corresponding to the UEP of the post-fault operation. Based on the CCA, the CCT is also analytically derived.

3) The PSC-VSC is able to re-synchronize with the power grid after around one cycle of oscillation, even if the fault clearing angle (time) is beyond the CCA (CCT). This self-restoration property reduces the risk of system collapse caused by the delayed fault clearance.

The conclusions of the paper are not limited to the PSC-VSC, but are equally applicable to any other control schemes with the first-order synchronization loop.

## APPENDIX

The detailed solution of (8) is provided as follows:
Defining

$$u = \tan\frac{\delta}{2} - \frac{b}{a} \quad (13)$$

Then

$$d\delta = \frac{2du}{1+(u+b/a)^2}, \quad \sin\delta = \frac{2(u+b/a)}{1+(u+b/a)^2} \quad (14)$$

Substituting (14) into (8), yields

$$t = \int \frac{2du}{a\left[u^2 + 1 - \frac{b^2}{a^2}\right]} \quad (15)$$

which leads to

$$t = \frac{2}{\sqrt{a^2-b^2}}\arctan\frac{u}{\sqrt{1-(b/a)^2}} + C \quad (16)$$

Substituting (13) into (16), the final solution of (8) is yielded, which is given by (9).

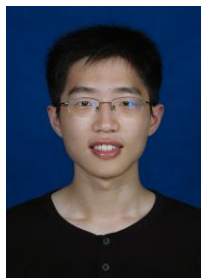

**Heng Wu** (S'17) received the B.S. degree in electrical engineering and automation and the M.S. degree in power electronic engineering both from Nanjing University of Aeronautics and Astronautics (NUAA), Nanjing, China, in 2012 and 2015, respectively. From 2015 to 2017, He was an Electrical Engineer with NR Electric Co., Ltd, Nanjing, China. He is currently working toward the Ph.D. degree in power electronic engineering in Aalborg University, Aalborg, Denmark. His research interests include the modelling and stability analysis of the power electronic based power systems.

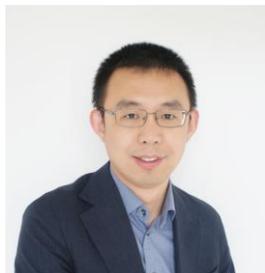

**Xiongfei Wang** (S'10-M'13-SM'17) received the B.S. degree from Yanshan University, Qinhuangdao, China, in 2006, the M.S. degree from Harbin Institute of Technology, Harbin, China, in 2008, both in electrical engineering, and the Ph.D. degree in energy technology from Aalborg University, Aalborg, Denmark, in 2013. Since 2009, he has been with the Aalborg University, Aalborg, Denmark, where he is currently an Associate Professor in the Department of Energy Technology. His research interests include modeling and control of grid-connected converters, harmonics analysis and control, passive and active filters, stability of power electronic based power systems.

Dr. Wang serves as an Associate Editor for the IEEE Transactions on Power Electronics, the IEEE Transactions on Industry Applications, and the IEEE Journal of Emerging and Selected Topics in Power Electronics. He is also the Guest Editor for the Special Issue "Grid-Connected Power Electronics Systems: Stability, Power Quality, and Protection" in the IEEE Transactions on Industry Applications. He received four IEEE prize paper awards, the outstanding reviewer award of IEEE Transactions on Power Electronics in 2017, and the IEEE PELS Richard M. Bass Outstanding Young Power Electronics Engineer Award in 2018.